\documentstyle[preprint,prd,eqsecnum,aps,epsf]{revtex}

\newcounter{num}
\newcommand{\be}{\begin{equation}}
\newcommand{\ene}{\end{equation}}
\newcommand{\bea}{\begin{eqnarray}}
\newcommand{\enea}{\end{eqnarray}}

\begin{document}
\vspace{1.cm}

\begin{center}
{\Large\bf Lepton Flavor Violation in Supersymmetric SO(10) 
Grand Unified Models}
\end{center}
\vspace{0.5cm}

\begin{center}
{ Xiao-Jun Bi, Yuan-Ben Dai, and Xiao-Yuan Qi}\\\vspace{3mm}
{\it Institute of Theoretical Physics,
 Academia Sinica, \\ P.O. Box 2735, Beijing 100080, People's Republic of China }
\end{center}
\vspace{1.5cm}

\begin{abstract}
The study for lepton flavor violation combined with the neutrino oscillation
may provide more information about the lepton
flavor structure of the grand unified theory. 
In this paper, we study two lepton flavor violation processes
, $\tau\rightarrow \mu\gamma$ and $Z\rightarrow \tau\mu$, in the context of
supersymmetric SO(10) grand unified models. We find the two processes are both
of phenomenological interest. In particular the latter may be important in some
supersymmetric parameter space where the former is suppressed. Thus,
Z--dacay may offer another chance for looking for lepton flavor violation.
\end{abstract}

\section {INTRODUCTION}
Super-Kamiokande data on the atmospheric neutrino anomaly presents a 
strong evidence for the existence of
neutrino oscillations. The anomaly can be explained by $\nu_\mu -\nu_\tau$ oscillation
with $\delta m_{23}^2=(1-8)\times 10^{-3}eV^2$ and a large
mixing angle $\sin^2 (2\theta_{\mu\tau})=0.8-1$ \cite{sk,skn}.
In addition, the long standing solar neutrino deficit 
can also be interpreted
as another type of neutrino oscillations \cite{solar,solarn}. 
Assuming that the LSND anomaly \cite{lsnd} will finally disappear all these 
observations about neutrinos can
be accommodated in a model with three very light active left--handed neutrinos.
Many such models have been proposed to explain the measured neutrino parameters 
since Super-Kamiokande first published the data
\cite{model}. Among them a natural explanation of the neutrino masses is 
provided by grand unified models which interpret the very light neutrino
masses compared with the quarks and charged leptons by the see--saw mechanism
\cite{ss}.
Especially in SO(10) grand unified models, in which the right--handed (RH) 
neutrinos
have masses of the order about unification scale and the lepton masses and 
the quark masses
are related, correct neutrino masses can be 
obtained \cite{mass}. The measured neutrino masses
are even regarded as a new evidence supporting the grand unification 
idea \cite{pre}. 
As been argued in Ref. \cite{mass} a value for $m(\nu_\tau)\sim\frac{1}{20}eV$ 
falls into a natural range predicted by a grand unified model based  on either
a string-unified G(224) model or a SO(10) grand unified model.
Furthermore, grand unified theory
(GUT) has the advantage that it relates the neutrino problem, its masses
and mixings, with charged lepton and quark masses and mixings into a large
fermion flavor problem and thus gives more definite predictions.

Conversely, the parameters measured in the neutrino sector can provide a window
on the grand unified theory study and they also give important feedback on the
problems of quarks and charged lepton masses as they are all related in the GUT. A 
direct inference of neutrinos being massive is the existence of 
a Kabayashi-Maskawa like matrix for the lepton sector. 
However, the processes violating lepton flavors due to this matrix 
is too small to be observed because of the tiny neutrino masses.
In the supersymmetric GUTs 
the high energy lepton flavor violation (LFV) interactions may leave
trace in the mass matrices of scalar partners of leptons by renormalization 
effects, 
which drives low energy LFV processes since the mass matrix of the charged 
leptons and that of sleptons can not be diagonalized simultaneously.
Therefore, the study of the LFV process
may provide important
information on the flavor structure of the supersymmetric GUTs. 
Various LFV processes
have been studied in different theoretical frameworks by a number of authors
\cite{ellis,gom,hasino,rges,other}.

There are two purposes of this work. One is to study
the LFV processes in the context of 
supersymmetric SO(10) grand unified models with a ``lopsided'' texture for
the mass matrices of the down quark and charged lepton which has been
advocated by a number of authors to accommodate the large $\nu_\mu -\nu_\tau$
mixing and small $V_{cb}$
\cite{barr,alt,mass}. If the experimental 
sensitivity on 
$\tau\rightarrow\mu\gamma$ can reach down to $1\times 10^{-9}$ \cite{ellis},
not only this process can be observed but it can even
be used to discriminate different fermion textures in the grand unified models.
Another purpose of this work
is to present the analytic formulas for the branching ratio of the process 
$Z\rightarrow\tau\mu$ in the minimal supersymmetric standard model (MSSM)
and the numerical study in the class of models considered by us. 
This calculation seems absent in the literature. This calculation is 
triggered by the Giga-Z option of the Tesla project which may expect the 
$10^9$ Z bosons at resonance \cite{giga}. The upper limit for the branching 
ratios could be improved down to $BR(Z\rightarrow\mu^\pm\tau^\mp) < f\times
2.2\times 10^{-8}$ with $f=0.2\sim 1$. We find this process is quite interesting
phenomenologically since in some supersymmetric parameter space $\tau\rightarrow\mu
\gamma$ being suppressed, this process can be important however.

The paper is arranged as follows. In sec II we discuss the origin of the 
LFV interactions
in a supersymmetric grand unified model and the Renormalization Group 
Equations(RGE). In sec III we present the formulas for the branching ratios of 
the two
processes. In Sec IV we briefly introduce a grand unified model proposed 
in Ref \cite{barr} that we used
in our calculations. The numerical results are presented in Sec V 
and we summarize and give conclusions in Sec VI. 

\section{Renormalization group equations and low energy supersymmetric
spectrum}
\subsection{Origin of Lepton Flavor Violation}
If the SM is extended with massive and non-degenarate neutrinos, 
LFV processes may be induced. However, such processes are highly suppressed due to 
the smallness of the neutrino masses. The branching ratio is proportional to
${\delta m^2_\nu}/{M^2_W}$ which is hopeless to be observed \cite{early}. 
When supersymmetry
enters the theory the scene changes completely. The LFV may also be induced 
through the generation mixing of the supersymmetric soft breaking terms of the 
lepton sector. However, arbitrary mixing
of these soft terms in the MSSM are not predictive. 
In our calculations we adopt the supergravity
mediated supersymmetry breaking mechanism to produce universal soft terms at
the GUT scale $M_{GUT}$, because non--universal soft terms at $M_{GUT}$
may produce too large low energy LFV observable effects \cite{ellis,gom}. 
The tree level 
universal soft terms may induce non-diagonal terms at low energy by radiative 
corrections including LFV interactions at high scale. Our procedure includes
calculating
the low energy supersymmetric soft terms which is not generation universal now by
integrating the RGEs and then calculate
the LFV branching ratios induced by these non-universal soft terms.

For a supersymmetric SO(10) grand unified model, the structure below $M_{GUT}$
where the grand unification has been 
spontaneously broken is the same as the MSSM
supplemented with MSSM singlet RH neutrino superfields. The superpotential of 
the lepton sector is now
\be
W=f_\nu^{ij} \hat{H}_2\hat{L}_i\hat{N}_j+f_l^{ij}\hat{H}_1\hat{L}_i\hat{E}_j
+\frac{1}{2}M^{ij}\hat{N}_i\hat{N}_j+\mu \hat{H}_1\hat{H}_2
\ene
where $f_\nu$ and $f_l$ are the Yukawa coupling matrices, $M$ is the RH neutrino
mass matrix. $i$, $j$ are the generation indices. Anti-symmetric tensor 
$\epsilon^{ab}$ is implict to contract the SU(2) doublets with $\epsilon^{12}=-1$.
In general, $f_\nu$ and $f_l$
can not be diagonalized simultaneously, which is the origin of LFV interactions.
Diagonalize $f_\nu$ and $f_l$ by bi-unitary rotations,
\bea
f^\delta_l&=&U_L^\dagger f_l U_R\nonumber\\
f^\delta_\nu&=&V_L^\dagger f_\nu V_R
\enea
where $U_{L,R}$, $V_{L,R}$ are all unitary matrices. Then define 
\be V_D=U_L^\dagger V_L
\ene
which is analog to the KM matrix $V_{KM}$ in the quark sector. 
$V_D$ is crucial for LFV processes. The RH neutrino masses are
not much lower in order compared to $M_{GUT}$ in SO(10) grand unified models. 
After the RH neutrinos are decoupled and $H_2$ gets 
VEV $v_2$ of weak scale we get three light left--handed (LH) Majorana neutrinos
with mass matrix $m_\nu=-(f_\nu v_2)M^{-1} (f_\nu v_2)^T$ by see--saw mechanism.
Suppose $m_\nu m_\nu^\dagger$
is diagonalized by $V^\nu_L$, then the matrix
\be 
V^{MNS}=U_L^\dagger V^\nu_L\ene
 determines the 
neutrino oscillation parameters. In the published grand unified models which 
emphasize the neutrino oscillations the large $\nu_\mu-\nu_\tau$ mixing angle
in $V^{MNS}$ is mainly coming from $U_L^\dagger$ \cite{model}. Thus, on the one hand, 
we may expect large rates for LFV processes due to large $\mu-\tau$ mixing
in $U_L^\dagger$, which at the same time causes large $\nu_\mu-\nu_\tau$ 
mixing in $V^{MNS}$ for neutrino oscillations, and, on the other hand, 
difference between $V_D$ and $V^{MNS}$ may be found by neutrino oscillations
and LFV processes. If so, important information about GUT structure may be
derived then.

The soft breaking terms for the lepton sector is
\cite{rosiek},
\bea
{\cal{L}}_{soft}&=&-m_{H_1}^2H_1^\dagger H_1-m_{H_2}^2H_2^\dagger
H_2
-(m_{\tilde{L}}^2)^{ij}\tilde{L}_i^\dagger\tilde{L}_j
-(m_{\tilde{R}}^2)^{ij}\tilde{R}_i^*\tilde{R}_j-(m_{\tilde{\nu}}^2)^{ij}
\tilde{\nu}_i^*\tilde{\nu}_j\nonumber\\
&&+\left(B\mu H_1H_2+\frac{1}{2}BM^{ij}\tilde{\nu}_i^*\tilde{\nu}_j^*
+(A_Ef_e)^{ij}H_1\tilde{L}_i\tilde{R}_j\right.\nonumber\\
&&\left.+(A_\nu f_\nu)^{ij}H_2\tilde{L}_i\tilde{\nu}_j+h.c.\right)
\enea
where $i$, $j$ are generations indices. 
At $M_{GUT}$ we assume the universal conditions,
\bea
m_{H_1}^2&=&m_{H_1}^2=m_0^2\\
m_{\tilde{L}}^2&=&m_{\tilde{R}}^2=m_{\tilde{\nu}}^2=m_0^2\\
A_E&=&A_\nu=A_0
\enea

Fig. 1 gives the explanation of the occurrence of low energy LFV processes.
The non-diagonal Yakawa couplings induce $\tilde{\nu}_\mu-\tilde{\nu}_\tau$
mixing through loop effects. This high energy process can be running down
by integrating the RGEs to low energy. Thus the study of LFV processes induced
by non-universal soft terms can actually reveal high energy fermion flavor
struture.
In the basis
where the $f_l$ is diagonal we can get the non-diaganol scalar mass 
in the first order of  approximation as
\bea
\left(\delta \tilde{m}^2\right)_{23}&\approx &\frac{1}{8\pi^2}f_{\nu}f_{\nu}^\dagger (3+a^2)m_0^2\log
\frac{M_{GUT}}{M_R}\nonumber\\
&\approx&\frac{1}{8\pi^2}(V_D)_{23}(V_D)_{33}\cdot f_{\nu_3}^2(3+a^2)m_0^2\log\frac{M_{GUT}}{M_R}
\label{dm}
\enea
where in the diagonalized Yukawa matrix $f_\nu^\delta$ only the (3,3) element
$f_{\nu_3}$ is kept.
$a$ is the universal parameter $A_0=a m_0$ and $M_R$ is the scale where
the RH neutrinos are decoupled.  

\subsection{RGEs running}

The RGEs used by us are given in the Appendix A. We have paid much effort to
keep the RGEs for the soft terms of lepton sector, 
which is relevant to our calculations, as complete as possible.
Specially speaking, both the diagonal and non-diagonal terms
are kept in the RGEs. However, only diagonal terms of the scalar
quark soft terms are kept because these non-diagonal terms are small and
not relevant to our calculations. For the Yukawa sector, 
only the (3,3) elements of the diagonalized Yukawa matrices 
$f_t$, $f_b$, $f_\tau$ and $f_{\nu_3}$ are kept. The evolution of the
mixing matrix $V_D$ from $M_{GUT}$ to $M_R$ is ignored. 

The integration procedure consists of iterative runnings of the RGEs from the GUT
scale to the low energy scale $M_Z$ and back for every set of inputs of $m_0$, 
$m_\frac{1}{2}$, $A_0$, $\tan\beta$ and the sign of $\mu$, which are the 
universal scalar masses, gaugino masses, scalar trilinear parameter, and 
the standard
vacuum expectation values ratio of the two Higgs fields and the Higgsino mass 
parameter,
until the low energy gauge couplings and the Yukawa couplings are all correct 
within a 
given range. The parameters $\mu$ and $B$ are given by low energy spontaneous 
breaking conditions\cite{rosiek,bar2}.

The RGEs at Appendix A is given in the basis where $f_\nu$ is diagonal.
 The RH neutrinos are decoupled at $M_R$. Below
$M_R$ the basis is rotated to the basis
where $f_l$ is diagonal. The MSSM RGEs are obtained by simply dropping the 
terms including $f_\nu$ and setting $V_D$ equal to 1. The basis rotation leads to
\bea
(m_{\tilde{L}}^2)_{below}&=&V_D (m_{\tilde{L}}^2)_{up} V_D^\dagger\nonumber\\
(A_E)_{below}&=&V_D (A_E) V_D^\dagger\ \ .
\enea
Below $M_{SUSY}=250GeV$ 
the SM beta functions are used \cite{barger}. Threshold effects
are taking into account by decoupling the corresponding particles at its running
mass $Q=m(Q)$.

Taking $\tan\beta=2\sim 10$, we show the numerical results as follows
\bea
\delta m_{\tilde{L}}^2&=&\left(\begin{array}{ccc}  0 & (0.92\sim 2.87)\times 10^{-3} & (-0.77\sim -2.06)\times 10^{-3}\\
   (0.92\sim 2.87)\times 10^{-3} & 0 & (0.97\sim 3.0)\times 10^{-2}\\
   (-0.77\sim -2.06)\times 10^{-3} & (0.97\sim 3.0)\times 10^{-2} & 0
   \end{array}\right)\ (3+a^2)m_0^2\\
 A_E&=&\left(\begin{array}{ccc} 
     0.7 & (2.42\sim 8.12)\times 10^{-4}  & (-2.41\sim -5.88)\times 10^{-4}\\
     (2.42\sim 8.12)\times 10^{-4} & 0.7 & (2.5\sim 8.67)\times 10^{-3}\\
    (-2.41\sim -5.88)\times 10^{-4} & (2.5\sim 8.67)\times 10^{-3} & 0.7
    \end{array}\right)\ \ A_0
\enea
where smaller $\tan\beta$ may give larger contribution.

\subsection{Low energy supersymmtric spectrum}
At low energy the supersymmetric particle masses and mixing angles
are obtained by diagonalizing the corresponding chargino, neutralino,
scalar neutrino and scalar lepton mass matrices numerically. The slepton
mass matrix is given by a $6\times 6$ matrix as
\be
m_{\tilde{l}}^2=\left( \begin{array}{cc} m_{LL}&m_{LR}\\
m_{LR}^\dagger&m_{RR}\end{array}\right)
\ene
where $m_{LL}$, $m_{LR}$, $m_{RR}$ are all $3\times 3$  matrices given by
\bea
(m_{LL})_{ij}&=&(m_{\tilde{L}}^2)_{ij}+(m_l^2)\delta_{ij}
+M_Z^2(-\frac{1}{2}+\sin^2\theta_W)
\cos 2\beta\delta_{ij}\ ,\nonumber\\
(m_{LR})_{ij}&=&-((A_E)^{ij}+\mu \tan\beta\delta^{ij})m_l^j\ ,\nonumber\\
(m_{RR})_{ij}&=&(m_{\tilde{R}}^2+m_l^2-M_Z^2\sin^2\theta_W\cos 2\beta)\delta_{ij}\ .
\enea
The $m_{\tilde{R}}^2$ is diagonal since only $f_l$ enters its RGE.
The full sneutrino mass matrix has a $12\times 12$ structure. However,
according to \cite{ellis,gom} 
if we only keep the first order of these terms perturbed by RH neutrino masses
we can have a very simple structure, which is relevant
to generation mixing,
\be
m_{\tilde{\nu}}^2=m_{\tilde{L}}^2+\frac{1}{2}M_Z^2\cos 2\beta\ \ .
\ene
The mass matrices of chargino and neutrino are standard and given at Appendix B.

\section{ Analytic formulas }
Fig. 2 gives the one loop diagrams relevant to the process 
$\tau\rightarrow\mu\gamma$. 
The amplitude of this process can be written in the general form
\be
M=em_i \bar{u}_j(p_2)i\sigma_{\mu\nu}p_3^\nu 
(A_LP_L+A_RP_R)u_i(p_1)\epsilon^\mu (p_3)\ \ ,
\ene
where $P_{L,R}=\frac{1}{2}(1\mp \gamma_5)$ are the chirality projection operators.
$i$, $j$ represent initial and final lepton flavor.
The most convenient way to calculate $A_L$ and $A_R$ is to pick up the one
loop momentum integral contributions which are proportional to 
$\bar{u}_j(p_2)P_{L,R}u_i(p_1)2p_1\cdot\epsilon$ respectively. 
The neutralino exchanging 
contribution is
\bea
A_L^{(n)}&=&\frac{1}{32\pi^2}(\frac{e}{\sqrt{2}\cos\theta})^2\frac{1}
{m_{\tilde{l}_\alpha}^2}\left[{B^{j\alpha a}}^*B^{i\alpha a}\frac{
1-6k+3k^2+2k^3-6k^2\log k}{6(1-k)^4}\right.\nonumber\\
&&+\left.\frac{m_{\chi_a^0}}{m_i}{B^{j\alpha a}}^*
A^{i\alpha a}\frac{1-k^2+2k\log k}{(1-k)^3}\right] \ ,\\
A_R^{(n)}&=&A_L^{(n)}\ (B\leftrightarrow A)\ ,
\enea
where $k={m_{\chi_a^0}^2}/{m_{\tilde{l}_\alpha}^2}$. $A$ and $B$ are
the lepton--slepton--neutralino coupling vertices given in Appendix B.
The corresponding contribution coming from exchanging charginos are
\bea
A_L^{(c)}&=&-\frac{g_2^2}{32\pi^2}{Z_\nu^{i\alpha}}^*Z_\nu^{j\alpha}
\frac{1}{m_{\tilde{\nu}_\alpha}^2}\left[ Z^-_{2a}{Z^-_{2a}}^*
\frac{m_im_j}{2M_W^2\cos^2\beta}
\frac{2+3k-6k^2+k^3+6k\log k}{6(1-k)^4}\right.+\nonumber\\
&&\left.\frac{m_{\chi^-_a}}{\sqrt{2}M_W\cos\beta}Z^+_{1a}Z^-_{2a}
\frac{m_j}{m_i}
\frac{3-4k+k^2+2\log k}{(1-k)^3}\right]\ ,\\
A_R^{(c)}&=&-\frac{g_2^2}{32\pi^2}{Z_\nu^{i\alpha}}^*Z_\nu^{j\alpha}
\frac{1}{m_{\tilde{\nu}_\alpha}^2}\left[ Z^+_{1a}{Z^+_{1a}}^*\frac
{2+3k-6k^2+k^3+6k\log k}{6(1-k)^4}\right.+\nonumber\\
&&\left.\frac{m_{\chi^-_a}}{\sqrt{2}M_W\cos\beta}Z^+_{1a}Z^-_{2a}\frac
{3-4k+k^2+2\log k}{(1-k)^3}\right]\ \ ,
\enea
where $k={m_{\chi_a^-}^2}/{m_{\tilde{\nu}_\alpha}^2}$. Mixing matrices
$Z_\nu$, $Z^+$ and $Z^-$ are given in Appendix B.

The branching ratio is given by 
\be
BR(\tau\rightarrow\mu\gamma)=\frac{\alpha_{em}}{4\pi}m_i^5(|A_L|^2+|A_R|^2)/
\Gamma_\tau \ \ ,
\ene
where $\Gamma_\tau=2.265\times 10^{-12}GeV$.

Fig. 3 gives the diagrams contributing to the $Z\rightarrow \tau\mu$ process.
Neglecting masses of the final fermions the amplitude is given as 
$M=\Sigma_ia_iM_i$, where $M_i$ are
\bea
M_{1,2}&=&\bar{u}(p_2)\gamma^\mu P_{L,R}v(p_1)\epsilon_\mu\ \ ,\\
M_{3,4}&=&\bar{u}(p_2)P_{L,R}v(p_1)p_{1}\cdot \epsilon\ \ 
\enea
respectively.
The corresponding analytic expressions of $a_i$ are given in appendix C.

The branching ratio for $Z\rightarrow \tau\mu$ is given by 
$Br(Z\rightarrow\mu\tau)={\Gamma(Z
\rightarrow \tau^\pm\mu^\mp)}/{\Gamma_Z}$, where $\Gamma_Z=2.49GeV$ and 
\bea
\Gamma(Z\rightarrow \mu\tau)&=&\frac{1}{48\pi M_Z}\cdot \sum |M|^2\nonumber\\
&=&\frac{1}{48\pi M_Z}\cdot \left[(|a_1|^2+|a_2|^2)\cdot 2M_Z^2+
(|a_3|^2+|a_4|^2)\cdot \frac{M_Z^4}{4}\right]\ \ ,
\enea
where $\Sigma$ represents the sum of $\mu$, $\tau$ spins and $Z$
polarizations.

\section{A SO(10) grand unified model}

To give definite predictions we will work within a specific model.
Before we turn to introduce this model it is worth noting that our calculations
are not very model sensitive. That is the reason why we concentrate on the 2--3
generation mixing. The mixing between the first two generations may be very 
sensitive to
different models and we will discuss the processes separately.
From Fig.1, we know that $\tau -\mu$ mixing 
is mainly determined by the factor
${\delta \tilde{m}^2_{23}}/{m_0^2}$, 
which, from Eq. (~\ref{dm}), depends only on the parameters
$(V_D)_{23}\cdot (V_D)_{33}$, $f_{\nu_3}$, $M_{GUT}$ and $M_R$. 
$M_{GUT}\approx 2\times 10^{16}GeV$
is required by any supersymmetric grand unified model. 
In the context of SO(10) grand unified
models $f_{\nu_3}$ is related to the top Yukawa coupling $f_t$ at the GUT scale.
In addition, with the SuperK data of $m_{\nu_\tau}\approx\frac{1}{20}eV$ 
$M_R\approx 2\times 10^{14}GeV$ 
is roughly determined by the see--saw mechanism. Thus the only model sensitive
parameter in fermion textures is $(V_D)_{23}$ ( $(V_D)_{33}$ is determined by
unitary condition of the $V_D$ matrix ).
In most published SO(10) models the large $\nu_\mu -\nu_\tau$ mixing is not
produced dominantly by $M_R$ mass matrix and thus produce large $\mu -\tau$
mixing, for example,
we have $(V_D)_{23}\cdot (V_D)_{33}
\sim (0.5-0.3)$ corresponding to $\theta_{23}\sim (20^\circ -70^\circ )$.
Thus our discussions on these branching ratios are useful for estimating 
the predictions for such a class of unified models, although they may be 
different in details.

We did our calculations within the model given by Albright {\it et al} \cite{barr}.
The model gives excellent predictions of quark and lepton masses and naturally
explains the largeness of $\nu_\mu -\nu_\tau$ mixing and the smallness of $V_{cb}$.
At $M_{GUT}$, after the SO(10) breaking to the MSSM, the fermion mass textures 
are given by
\bea
U&=&\left(\begin{array}{ccc}\eta & 0 & 0\\
			  0 & 0 & -\epsilon/3\\
			  0 & \epsilon/3 & 1 \end{array}\right)\ M_U\ \ ,
\ \ D=\left(\begin{array}{ccc}0 & \delta & \delta ^{'}e^{i\phi}\\
			  \delta & 0 & -\epsilon/3\\
			  \delta ^{'}e^{i\phi} & \sigma+\epsilon/3 & 1 \end{array}\right)\ M_D\ \ ,\\
N&=&\left(\begin{array}{ccc}\eta & 0 & 0\\
			  0 & 0 & \epsilon\\
			  0 & -\epsilon & 1 \end{array}\right)\ M_U\ \ ,
\ \ \ \ \ \ \ \ L=\left(\begin{array}{ccc}0 & \delta & \delta ^{'}e^{i\phi}\\
			  \delta & 0 & \sigma+\epsilon\\
			  \delta ^{'}e^{i\phi} & -\epsilon & 1 \end{array}\right)\ M_D\ \ ,
\enea
where $U$, $D$, $N$ and $L$ are the up quark, down quark, neutrino and lepton
mass matrices respectively. The most remarkable feature of the textures is 
the lopsided parameter $\sigma\sim 1$ at $L$ and $D$. According to the SU(5)
relation $D=L^T$, the large $\nu_\mu-\nu_\tau$ mixing due to the $\sigma$ term
in $L$ is related to the right--handed down quark mixing, which has no observable
physical effects. The smallness of quark mixing $V_{cb}$ is determined by the 
parameter $\epsilon\sim 0.1$, which translates to the right--handed lepton mixing.

The model predicts the lepton sector 2-3 mixing with 
$\theta_{23}\sim 63^\circ$, which leads
to the only model structure sensitive quantity for our processes
$(V_D)_{23}\cdot (V_D)_{33}\sim 0.4$. 

Taking all the fermion masses and $V_{KM}$ elements at $M_Z$ given
in Ref. \cite{fus} as inputs, 
we calculate the corresponding values at $M_{GUT}$ with several values of
$\tan\beta$ by integrating the RGEs
and fit the parameters in Eq. (4.1) and (4.2). We find these parameters are
not sensitive to the supersymmetric parameters, except that $M_U$ becomes larger
as taking small $\tan\beta$. To keep the predicted neutrino masses in the
correct range we take $M_R=5\times 10^{14}GeV$ when $\tan\beta=2$ and
$M_R=2\times 10^{14}GeV$ when $\tan\beta=5,\ 10$ in our later calculations.

\section{ Numerical result and discussion }


The relevant parameters on predicting the branching ratios 
are the universal supersymmetry soft breaking
parameters, $m_0$, $m_\frac{1}{2}$, $A_0$, $\tan\beta$ and the lepton sector 
mixing matrix $V_D$ defined in Eq. (2.3). 
We will present dependence of the branching ratio of the processes 
$\tau\rightarrow\mu\gamma$ and $Z\rightarrow\tau\mu$ on
the soft parameters in the model of \cite{barr}. 
In our calculations the
soft parameters are constrained by various theoretical and experimental
considerations \cite{jet}, such as the LSP of the model should be a neutralino
and masses of all 
supersymmetric particles be beyond present mass limits and so on.

In Figs. 4$\sim$7, we plot the branching ratio of $\tau\rightarrow\mu\gamma$ as functions
of $m_0$ for several sets of other soft parameters. 
Two experimental bounds are plotted
in every figure, which correspond to the present experimental limit 
$BR(\tau\rightarrow\mu\gamma)\ <\ 1.1\times 10^{-6}$ \cite{pdg} and 
the expected sensitivity of future experiments
$BR(\tau\rightarrow\mu\gamma)\ <\ 1.0\times 10^{-9}$\cite{ellis}.
The general trend of 
$BR(\tau\rightarrow\mu\gamma)$ is decreasing dramatically as $m_0$ increases. 
In Fig. 4, the branching ratio is plotted for $m_\frac{1}{2}=100GeV$,
$\tan\beta=5$ and $A_0=2m_0,\ m_0,\ 0$, taking $\mu$ either positive or negative.
That branching ratio decreases with $a=A_0/m_0$ is easily understood
due to the Eq. (1). It is interesting to note that although this process
can not be observed at present, it will be detectable in the future experiments
in a large part of the soft parameter space. 

Fig. 5 presents the branching ratio for $m_\frac{1}{2}=100,\ 200,\ 400GeV$
and $A_0=m_0$, $\tan\beta=5$. We can find that the branching ratio is also very
sensitive to the parameter $m_\frac{1}{2}$. It will decrease quickly with
increasing $m_\frac{1}{2}$. If $m_\frac{1}{2}\ >\ 400GeV$, this process will
be non-observable as displayed in the figure. 
In Fig. 6, we plot the branching ratio for 
$m_\frac{1}{2}=100GeV$, $\tan\beta=10$ and $A_0=\pm 2m_0,\ \pm m_0,\ 0$. 
We find that the sign of $A_0$ is not very important in the order of magnitude
 estimate.
However, the sign of $A_0$ is still relevant to the precise predictions.
Fig. 7 plots the branching ratio at 
different $\tan\beta$. The branching ratio increases as $\tan\beta$ becomes
large. It is explained in Ref. \cite{hasino} that the dominant term 
of the amplitude for $\tau\rightarrow\mu\gamma$ is proportional to 
$\tan\beta$. We note that at $\tan\beta=2$
the sign of $\mu$ can be significant. The branching ratio when $\mu$ is
negative is much smaller than that when $\mu$ is positive.

In Figs. 8$\sim$10, we present the quantitative results of the 
branching ratio of the decay $Z\rightarrow\tau\mu$.
We find in most parameter space this process can not be detected by
the Giga-Z option \cite{giga}. However, we can still find several interesting
features in this process which are different from the $\tau$ 
radiative decay process. The most remarkable feature about this LFV process 
is that its branching ratio becomes large at first and then approaches to a
constant when $m_0$ is increasing, which
gives a sharp contrast with the process $\tau\rightarrow\mu\gamma$. Thus
this process may have advantage
over the $\tau\rightarrow\mu\gamma$ process in some region of parameter space. 
The reason for this different behavior between the two processes can be traced
back to the different coupling structures between the processes as shown in 
Eqs. (3.1) and (3.7), (3.8). The magnetic structure of $\tau$ decay determines 
that its amplitude is inversely proportional to the sfermion masses square, 
whereas there
is a vector current coupling in Eq. (3.7) which determines the $Z\rightarrow\mu\tau$
amplitude's trend as increasing $m_0$.

Fig. 8 shows the branching ratio as a
function of $m_0$ for $m_\frac{1}{2}=100GeV$, $\tan\beta=5$ and $A_0=2m_0,\
m_0,\ 0$ with both positive and negative sign of $\mu$. 
The sign of $\mu$ is quite irrelevant when $m_0\ >\ 500GeV$. 
Fig. 9 displays the same function for different $m_\frac{1}{2}$
values, $100,\ 200,\ 400GeV$, with $A_0=2m_0$ and $\tan\beta=5$. The dependence
of $BR(Z\rightarrow\tau\mu)$ on $a$ and $m_\frac{1}{2}$ is similar to that
of $BR(\tau\rightarrow\mu\gamma)$. We note again that 
the sign of $\mu$ is irrelevant
when $m_{\frac{1}{2}}\ >\ 200GeV$.
Fig. 10 gives the branching
ratio for $\tan\beta=2,\ 5,\ 10$ with $A_0=2.5m_0$, $m_\frac{1}{2}=150GeV$.
We can see another important feature of this LFV process that it is more 
favorable for the
small $\tan\beta$ value in contrast with the $\tau\rightarrow\mu\gamma$ process.
At the extreme case, the branching ratio may access $1\times 10^{-8}$, which
may be detectable. The relationship between the branching ratio and $\tan\beta$
is easily understood. For small $\tan\beta$ large non--universal soft mass
term $\delta \tilde{m}_{23}^2$ will be produced due to a large $f_{\nu_3}$ in 
Eq. (~\ref{dm}). We note that when
$\tan\beta$ becomes smaller, the $BR(Z\rightarrow\tau\mu)$ increases quickly.

In summary, according to our calculations 
we find that the $\tau\rightarrow\mu\gamma$ is more feasible
than $Z\rightarrow\tau\mu$ to reveal charged lepton flavor violation in the
context of SO(10) SUSY--GUTs. In most parameter space $\tau\rightarrow\mu\gamma$
has a branching ratio that can be detected in the future experiments \cite{ellis}.
The $Z\rightarrow\tau\mu$ is hopeful to be detected in a small parameter space.
We note the remarkable feature of the $Z$ decay process that its dependence on
the supersymmetry soft parameters $\tan\beta$ and $m_0$ is opposite to that
of $\tau$ decay. Thus it can supplement the LFV search besides the 
$\tau\rightarrow\mu\gamma$ process.

\section{Summary and conclusion}

In this work, we present the dependence of the branching ratios of 
two charged lepton flavor violation processes $\tau\rightarrow\mu\gamma$
and $Z\rightarrow\tau\mu$ on the supersymmetry soft breaking parameters
in the context of SO(10) grand unified models with the "lopsided" texture of
mass matrices for charged leptons and down quarks. We expect these processes
may be detected in the future experiments. The first process is more
hopeful to be observed. The second process may offer useful information
about the soft parameters if it is also observed. The different behaviors
of the two processes depending on $\tan\beta$ and $m_0$ implies that
the simultaneous study of the two processes will be interesting.

We expect the study of charged lepton flavor process may provide
another window of high energy physics besides the neutrino oscillation study.
The combined study of the LFV processes and neutrino oscillation
may feed light on the sector of right-handed neutrinos, which may
be necessary in a model to naturally explain the small neutrino masses.

\section*{ACKNOWLEDGMENTS}
Y-B. Dai's work is supported by the National Science Foundation of China.

\begin{center}
{\bf Appendix A}\\
\end{center}
\setcounter{num}{1}
\setcounter{equation}{0}
\def\theequation{\Alph{num}.\arabic{equation}}
\newcommand{\fa}{\frac{1}{16\pi^2}}
In this appendix we give the renormalization group equations
of the MSSM suplemented with RH neutrinos\cite{hasino,rges}. 
The two--loop RGEs of 
the gauge couplings can be found in many literatures which will not
be affected by the presence of the RH neutrinos. We give one--loop RGEs
of the Yakawa coupling matrices and the soft terms which are affected by the
presence of RH neutinos. In this appendix we denote the Yukawa
couplings of up quark, down quark, lepton and neutrino as $U$, $D$, $E$
and $N$ respectively. Denote the soft terms as $A_U\cdot U=U_A$ and so on.
The RGEs of the Yukawa couplings are
\bea
\frac{dU}{dt}&=&\frac{1}{16\pi^2}\left[-\Sigma c_ig_i^2+3UU^\dagger+
DD^\dagger+Tr[3UU^\dagger+NN^\dagger]\right]U\ \ ,\\
\frac{dD}{dt}&=&\frac{1}{16\pi^2}\left[-\Sigma c_i^{'}g_i^2+3DD^\dagger+
UU^\dagger+Tr[3DD^\dagger+EE^\dagger]\right]D\ \ ,\\
\frac{dE}{dt}&=&\frac{1}{16\pi^2}\left[-\Sigma c_i^{''}g_i^2+3EE^\dagger+
NN^\dagger+Tr[3DD^\dagger+EE^\dagger]\right]E\ \ ,\\
\frac{dN}{dt}&=&\frac{1}{16\pi^2}\left[-\Sigma c_i^{'''}g_i^2+3NN^\dagger+
EE^\dagger+Tr[3UU^\dagger+NN^\dagger]\right]N\ \ ,
\enea
where $t=\log Q$, $c_i=(\frac{13}{15},3,\frac{16}{3})$, 
$c_i^{'}=(\frac{7}{15},3,\frac{16}{3})$, $c_i^{''}=(\frac{9}{5},3,0)$, 
$c_i^{'''}=(\frac{3}{5},3,0)$.
The RGEs of $\mu$ and soft parameters of Higgs sector are
\bea
\frac{d\mu}{dt}&=&\frac{1}{16\pi^2}\left[-\Sigma c_i^{'''}g_i^2+
Tr[3UU^\dagger+3DD^\dagger+EE^\dagger+NN^\dagger]\right]\mu\ \ ,\\
\frac{dB}{dt}&=&\frac{2}{16\pi^2}\left[-\Sigma c_i^{'''}g_i^2M_i+
Tr[3UU_A+3DD_A+EE_A+NN_A]\right]\ \ ,\\
\frac{dm_{H_U}^2}{dt}&=&\frac{2}{16\pi^2}\left[-\Sigma c_i^{'''}g_i^2M_i^2+
3Tr[U(M_{Q_L}^2+M_{U_R}^2)U^\dagger+m_{H_U}^2UU^\dagger+U_AU_A^\dagger]\right.\nonumber\\
&&\left.+
Tr[M_{Q_L}^2NN^\dagger+N M_{\tilde{\nu}_R}^2N^\dagger
+m_{H_U}^2NN^\dagger+N_AN_A^\dagger]\right]\ \ ,\\
\frac{dm_{H_D}^2}{dt}&=&\frac{2}{16\pi^2}\left[-\Sigma c_i^{'''}g_i^2M_i^2+
3Tr[D(M_{Q_L}^2+M_{D_R}^2)D^\dagger+m_{H_D}^2DD^\dagger+D_AD_A^\dagger]\right.\nonumber\\
&&\left.+
Tr[E(M_{\tilde{L}}^2+ M_{\tilde{R}}^2)E^\dagger
+m_{H_D}^2EE^\dagger+E_AE_A^\dagger]\right]\ \ .
\enea
$M_i$ in the above expressions are the gaugino masses whose RGEs are same
as those in the MSSM.

Then we give RGEs of soft mass terms of lepton sector 
\bea
\frac{dM_{\tilde{L}}^2}{dt}&=&\frac{2}{16\pi^2}\left[-\Sigma c_i^{'''}g_i^2M_i^2+
\frac{1}{2}[NN^\dagger M_{\tilde{L}}^2+M_{\tilde{L}}^2NN^\dagger]+
\frac{1}{2}[EE^\dagger M_{\tilde{L}}^2+M_{\tilde{L}}^2EE^\dagger]\right.\nonumber\\
&&\left.+EM_{\tilde{R}}^2E^\dagger+m_{H_D}^2EE^\dagger+E_AE_A^\dagger+
N M_{\tilde{\nu}_R}^2N^\dagger+m_{H_U}^2NN^\dagger+N_AN_A^\dagger\right]\ ,\\
\frac{dM_{\tilde{R}}^2}{dt}&=&\frac{2}{16\pi^2}\left[-\frac{12}{5}g_1^2M_1^2+
E^\dagger EM_{\tilde{R}}^2+M_{\tilde{R}}^2E^\dagger E+
2(E^\dagger M_{\tilde{L}}^2 E+m_{H_D}^2E^\dagger E+E_A^\dagger E_A)\right]\ ,\\
\frac{dM_{\tilde{\nu}_R}^2}{dt}&=&\frac{2}{16\pi^2}\left[
N^\dagger N M_{\tilde{\nu}_R}^2+M_{\tilde{\nu}_R}^2 N^\dagger N +
2(N^\dagger  M_{\tilde{L}}^2 N+m_{H_U}^2 N^\dagger N+N_A^\dagger N_A)\right]\ .
\enea

The RGEs of trilinear terms of lepton sector are 
\bea
\frac{dA_E}{dt}&=&\frac{1}{16\pi^2}\left[-2\Sigma c_i^{''}g_i^2M_i+
2Tr(3A_DDD^\dagger+A_EEE^\dagger)+2A_N NN^\dagger\right.\nonumber\\
&&\left.+(5EE^\dagger+NN^\dagger)A_E+A_E(EE^\dagger-NN^\dagger)\right]\ ,\\
\frac{dA_N}{dt}&=&\frac{1}{16\pi^2}\left[-2\Sigma c_i^{'''}g_i^2M_i+
2Tr(3A_UUU^\dagger+A_NNN^\dagger)+2A_E EE^\dagger\right.\nonumber\\
&&\left.+(5NN^\dagger+EE^\dagger)A_N+A_N(N^\dagger-EE^\dagger)\right]\ .
\enea

In the basis where $N$ is diagonal and only keep the third family Yukawa
coupling constants $f_t$, $f_b$, $f_{\nu_3}$ and $f_\tau$, then the RGEs are
simplified as
\bea
\frac{df_t}{dt}&=&\fa\left[-\Sigma c_ig_i^2+6f_t^2+f_b^2+f_{\nu_3}^2\right]f_t\ ,\\
\frac{df_b}{dt}&=&\fa\left[-\Sigma c^{'}_ig_i^2+6f_b^2+f_t^2+f_{\tau}^2\right]f_b\ ,\\
\frac{df_{\nu_3}}{dt}&=&\fa\left[-\Sigma c^{'''}_ig_i^2+4f_{\nu_3}^2+3f_t^2+f_\tau^2|V_{33}|^2\right]f_{\nu_3}\ ,\\
\frac{df_\tau}{dt}&=&\fa\left[-\Sigma c^{''}_ig_i^2+4f_\tau^2+3f_b^2+f_{\nu_3}^2|V_{33}|^2\right]f_\tau\ ,\\
\frac{d\mu}{dt}&=&\fa\left[-\Sigma c^{'''}_ig_i^2+3f_t^2+3f_b^2+f_{\nu_3}^2+f_\tau^2\right]\mu\\
\frac{dB}{dt}&=&\frac{2}{16\pi^2}\left[-\Sigma c^{'''}_ig_i^2M_i+
3f_t^2A_t+3f_b^2A_b+f_{\nu_3}^2A_{\nu_3}+f_\tau^2A_\tau\right]\\
\frac{dm_{H_U}^2}{dt}&=&\frac{2}{16\pi^2}\left[-\Sigma c^{'''}_ig_i^2M_i^2+
3f_t^2(M_{\tilde{t}_L}^2+M_{\tilde{t}_R}^2+m_{H_U}^2+A_t^2)\right.\nonumber\\
&&+\left. f_{\nu_3}^2(M_{\tilde{\tau}_L}^2+M_{\tilde{\nu}_R}^2+m_{H_U}^2+A_{\nu_3}^2)\right]\ ,\\
\frac{dm_{H_D}^2}{dt}&=&\frac{2}{16\pi^2}\left[-\Sigma c^{'''}_ig_i^2M_i^2+
3f_b^2(M_{\tilde{t}_L}^2+M_{\tilde{b}_R}^2+m_{H_D}^2+A_b^2)\right.\nonumber\\
&&+\left. f_{\tau}^2(M_{\tilde{\tau}_L}^2+M_{\tilde{\tau}_R}^2+m_{H_D}^2+A_\tau^2)\right]\ ,\\
\left(\frac{M_{\tilde{L}}^2}{dt}\right)_{ij}&=&\frac{2}{16\pi^2}\left[
(-\Sigma c_i^{'''}g_i^2M_i^2)\delta_{ij}+V_{3i}^*V_{3j}f_\tau^2\left(\frac{1}{2}(
(M_{\tilde{L}}^2)_{ii}+(M_{\tilde{L}}^2)_{jj})+M_{\tilde{\tau}_R}^2+m_{H_D}^2+(A_E)_{ii}(A_E)_{jj}\right)\right.\nonumber\\
&&\left.+f_{\nu_3}^2\left( \frac{1}{2}((M_{\tilde{L}}^2)_{3j}\delta_{3i}
+(M_{\tilde{L}}^2)_{i3}\delta_{3j})+m_{\tilde{\nu}_3}^2\delta_{i3}\delta_{j3}+
M_{H_U}^2\delta_{i3}\delta_{j3}+(A_\nu)_{i3}(A_\nu)_{j3}\right)\right]\ ,\\
\left(\frac{dm_{\tilde{R}}}{dt}\right)_{ii}&=&\frac{2}{16\pi^2}\left[
-\frac{12}{5}g_1^2M_1^2+2f_\tau^2((M_{\tilde{L}}^2)_{ii}+(M_{\tilde{R}}^2)_{ii}
+m_{H_D}^2+A_\tau^2)\delta_{i3}\right]\ ,\\
\left(\frac{dm_{\tilde{\nu}_R}}{dt}\right)_{ii}&=&\frac{2}{16\pi^2}\left[
2f_{\nu_3}^2((M_{\tilde{L}}^2)_{ii}+(M_{\tilde{\nu}_R}^2)_{ii}
+m_{H_U}^2+A_{\nu}^2)\right]\delta_{i3}\ ,\\
\frac{dA_t}{dt}&=&\frac{2}{16\pi^2}\left[-\Sigma c_ig_i^2M_i+6f_t^2A_t+f_b^2A_b+f_{\nu_3}^2A_{\nu_3}\right]\ ,\\
\frac{dA_b}{dt}&=&\frac{2}{16\pi^2}\left[-\Sigma c^{'}_ig_i^2M_i+6f_b^2A_b+f_\tau^2A_\tau+f_t^2A_t\right]\ ,\\
\left(\frac{dA_E}{dt}\right)_{ij}&=&\fa\left[
(-2\Sigma c_i^{''}g_i^2M_i+6f_b^2A_b+2f_\tau^2A_\tau)\delta_{ij}+V_{3i}^*V_{3j}f_\tau^2(
5(A_E)_{jj}+(A_E)_{ii})+\right.\nonumber\\
&&\left.f_{\nu_3}^2(2(A_\nu)_{i3}\delta_{j3}+(A_E)_{3j}\delta_{i3}-(A_E)_{i3}\delta_{j3})
\right]\ ,\\
\left(\frac{dA_\nu}{dt}\right)_{ij}&=&\fa\left[
(-2\Sigma c_i^{'''}g_i^2M_i+6f_t^2A_t+2f_{\nu_3}^2A_{\nu_3})\delta_{ij}+V_{3i}^*V_{3j}f_\tau^2(
2(A_E)_{ii}+(A_\nu)_{ii}-(A_\nu)_{ii})+\right.\nonumber\\
&&\left.f_{\nu_3}^2(5(A_\nu)_{3j}\delta_{i3}+(A_\nu)_{i3}\delta_{j3})\right]\ .
\enea
The matrix $V$ in the above equations refers to $V_D$ defined in Eq. (2.3).
Below $M_R$, RH neutrinos are decoupled and the RGEs of the MSSM are used.
This can be achieved by setting $f_{\nu_3}=0$ and $V_{ij}=\delta_{ij}$ in the
above expressions.

\begin{center}
{\bf Appendix B}\\
\end{center}
\setcounter{num}{2}
\setcounter{equation}{0}
\def\theequation{\Alph{num}.\arabic{equation}}
In this appendix we list the relevant pieces of the Lagrangian and conventions
which we take in our calculations \cite{rosiek}.
The Lagrangian pieces are
\bea
{\cal L}_{\tilde{\nu}\tilde{\nu}Z}&=&\frac{-ig_2}{2\cos\theta_W}
( {\tilde{\nu}}^{I*}
\stackrel{\leftrightarrow}{\partial_\mu}\tilde{\nu}^J)Z^\mu\ ,\\
{\cal L}_{\tilde{l}\tilde{l}Z}&=&\frac{ig_2}{\cos\theta_W}
\left(\frac{1}{2}Z_L^{Ii}{Z_L^{Ij}}^*-\sin^2\theta_W\delta_{ij}\right)(
\tilde{l}_i^*\stackrel{\leftrightarrow}{\partial_\mu}\tilde{l}_j)Z^\mu\ ,\\
{\cal L}_{\chi^+\chi^-Z}&=&\frac{g_2}{2\cos\theta_W}\bar{\chi^-_j}\gamma^\mu[
{Z^+_{1j}}^*Z^+_{1i}P_R+Z^-_{1j}{Z^-_{1i}}^*P_L+\delta_{ij}\cos 2\theta_W]
\chi^-_i\ Z_\mu\ , \\
{\cal L}_{\chi^0\chi^0Z}&=&\frac{g_2}{2\cos\theta_W}\bar{\chi^0_i}\gamma^\mu\left[
({Z_N^{4i}}^*Z_N^{4j}-{Z_N^{3i}}^*Z_N^{3j})P_L-
(Z_N^{4i}{Z_N^{4j}}^*-Z_N^{3i}{Z_N^{3j}}^*)P_R\right]\chi^0_j\ Z_\mu\nonumber\\
&=&\frac{g_2}{2\cos\theta_W}\bar{\chi^0_i}\gamma^\mu\left[C^{ij}P_L+D^{ij}P_R\right]
\chi^0_j\ Z_\mu\ ,\\
{\cal L}_{\chi^+l\tilde{\nu}}&=&-g_2\bar{l^I}\left[{Z^+_{1i}}^*P_R-\frac{m_{l^I}}{
\sqrt{2}M_W\cos\beta}Z^-_{2i}P_L\right]Z_{\nu}^{IJ}\chi^-_i\tilde{\nu}_J+h.c.\\
{\cal L}_{\chi^0l\tilde{l}}&=&\frac{e}{\sqrt{2}\cos\theta_W}\bar{\chi^0_j}\left[
\left(Z_L^{Ii}(Z_N^{1j}+Z_N^{2j}\cot\theta_W)-\cot\theta_W\frac{m_{l^I}}{
M_W\cos\beta}Z_L^{(I+3)i}{Z_N^{3j}}\right)P_L\right.\nonumber\\
&&-\left.\left(2Z_L^{(I+3)i}{Z_N^{1j}}^*+\cot\theta_W
\frac{m_{l^I}}{M_W\cos\beta}Z_L^{Ii}{Z_N^{3j}}^*\right)P_R\right]l^I\
{\tilde{l}}_i^*+h.c.\nonumber\\
&&=\frac{e}{\sqrt{2}\cos\theta_W}\bar{\chi^0_j}\left[A^{Iij}P_L+B^{Iij}P_R\right]
l^I\tilde{l}_i^*+h.c.
\enea
The abbreviations defined in the above expressions will be used
in the next Appendix.

The mixing matrices $Z$ in the above expressions are given below.
The scalar lepton and scalar neutrino mass matrices are given in Sec II. The
corresponding mixing matrices are defined as
\bea
Z_L^\dagger m_{\tilde{l}}^2 Z_L&=&diag\left(m_{\tilde{l_i}}^2\right),\ \ i=1\ldots 6\\
Z_\nu^\dagger m_{\tilde{\nu}}^2 Z_\nu&=&diag\left( m_{\tilde{\nu_i}}^2\right),\ \ i=1,\ 2,\ 3
\enea
The mass matrix of charginos is
\begin{equation}
M_{\chi} = \left[\matrix{
{m}_2 & \sqrt{2} M_W \sin\beta \cr
\sqrt{2} M_W \cos\beta & \mu } \right] \ \ .
\end{equation}
The mixing matrices $Z^\pm$ is defined as
\begin{equation}
(Z^-)^TM_{\chi} Z^+
= {\rm diag}\left(m_{{\chi}_1}, m_{{\chi}_2}\right) \ .
\end{equation}
The mass matrix for neutralinos is
\begin{equation}
M_{{\chi}^0} = \left[\matrix{
{m}_1   & 0 & -M_Z\cos\beta\sin\theta_W & M_Z\sin\beta\sin\theta_W \cr
0 & {m}_2 & M_Z\cos\beta\cos\theta_W & -M_Z\sin\beta\cos\theta_W \cr
-M_Z\cos\beta\sin\theta_W & M_Z\cos\beta\cos\theta_W & 0 & -\mu\cr
M_Z\sin\beta\sin\theta_W & -M_Z\sin\beta\cos\theta_W & -\mu & 0 } \right] \ , \ \ 
\end{equation}
which is diagonalized by
\begin{equation}
Z_N^T M_{{\chi}^0} Z_N
= {\rm diag}\left(m_{{\chi}^0_1}, m_{{\chi}^0_2},
m_{{\chi}^0_3}, m_{{\chi}_4^0}\right) \ \ .
\end{equation}

\begin{center}
{\bf Appendix C}\\
\end{center}
\setcounter{num}{3}
\setcounter{equation}{0}
\def\theequation{\Alph{num}.\arabic{equation}}
In this appendix we give the analytic expressions of $a_i$ defined in Sec III.

$a_1$ is given by
\be
a_1=\frac{g_2}{2\cos\theta_W}\frac{g_2^2}{16\pi^2}
(a_1(a)+a_1(b)+a_1(c)+a_1(d)+2a_1(e)+2a_1(f))\ ,
\ene
where $a_1(i)$ is coming from the coresponding Feynman diagram. They are given by
\bea
a_1(a)&=&|Z^+_{1a}|^2Z_\nu^{i\alpha}{Z_\nu^{j\alpha}}^*(-2)C_{00}\ , \\
a_1(b)&=&-\tan^2\theta_W\left(\frac{1}{2}Z_L^{I\alpha}Z_L^{I\beta}-\sin^2\theta_W
\delta^{\alpha\beta}\right)(-2){A^{i\alpha a}}^*A^{j\beta a}C_{00}\ , \\
a_1(c)&=&-{Z^+_{1a}}^*Z^+_{1b}Z_\nu^{i\alpha}{Z_\nu^{j\alpha}}^*
\left[\left((Z^-_{1b}{Z^-_{1a}}^*+\delta^{ab}\cos 2\theta_W)m_{\chi^-_a}m_{\chi^-_b}
-({Z^+_{1b}}^*Z^+_{1a}+\delta^{ab}\cos 2\theta_W)m_{\tilde{\nu}_\alpha}^2\right)
C_0\right.\nonumber\\
&&\left.-({Z^+_{1b}}^*Z^+_{1a}+\delta^{ab}\cos 2\theta_W)(B_0-2C_{00})\right]\ , \\
a_1(d)&=&-\frac{1}{2}\tan^2\theta_W{A^{i\alpha a}}^*A^{j\alpha b}\left[
2D^{ab}C_{00}+\left( C^{ab}m_{\chi^0_a}m_{\chi^0_b}-D^{ab}m_{\tilde{l}_\alpha}^2
\right)C_0-D^{ab}B_0\right]\ , \\
a_1(e)&=&-2(0.5-\sin^2\theta_W)|Z^+_{1a}|^2Z_\nu^{i\alpha}{Z_\nu^{j\alpha}}^*(
B_0+B_1)\ , \\
a_1(f)&=&-\tan^2\theta_W(0.5-\sin^2\theta_W){A^{i\alpha a}}^*A^{j\alpha a}(B_0+B_1)\ ,
\enea
where $i$, $\alpha$ and $a$ represent the flavors of lepton, slepton or 
sneutrino and chargino or neutralino respectively.
$a_2$ is given by
\be
a_2=\frac{g_2}{2\cos\theta_W}\frac{g_2^2}{16\pi^2}(a_2(b)+a_2(d)+2a_2(f))\ ,
\ene
where
\bea
a_2(b)&=&-\tan^2\theta_W(\frac{1}{2}Z_L^{I\alpha}Z_L^{I\beta}-\sin^2\theta_W\delta^{\alpha\beta})
(-2){B^{i\alpha a}}^*B^{j\alpha a}C_{00}\ , \\
a_2(d)&=&-\frac{1}{2}\tan\theta_W{B^{i\alpha a}}^*B^{j\alpha b}\left[
2C^{ab}C_{00}+\left(D^{ab}m_{\chi^0_a}m_{\chi^0_b}-C^{ab}m_{\tilde{l}_\alpha}^2
\right)C_0-C^{ab}B_0\right]\ , \\
a_2(f)&=&\tan^2\theta_W\sin^2\theta_W{B^{i\alpha a}}^*B^{j\alpha a}(B_0+B_1)\ ,
\enea
Then we have
\be
a_i=\frac{g_2}{2\cos\theta_W}\frac{g_2^2}{16\pi^2}(a_i(b)+a_i(d)),\ 
i=3,\ 4\ene
and
\bea
a_3(b)&=&2\tan^2\theta_W(\frac{1}{2}Z_L^{I\alpha}Z_L^{I\beta}-\sin^2\theta_W\delta^{\alpha\beta})
{B^{i\alpha a}}^*A^{j\beta a}m_{\chi^0_a}(C_0+C_1+C_2)\ , \\
a_3(d)&=&\tan^2\theta_W{B^{i\alpha a}}^*A^{j\alpha b}(m_{\chi^0_b}C^{ab}C_1+
m_{\chi^0_a}D^{ab}C_2)\ , \\
a_4(b)&=&2\tan^2\theta_W(\frac{1}{2}Z_L^{I\alpha}Z_L^{I\beta}-\sin^2\theta_W\delta^{\alpha\beta})
{A^{i\alpha a}}^*B^{j\beta a}m_{\chi^0_a}(C_0+C_1+C_2)\ , \\
a_4(d)&=&\tan^2\theta_W{A^{i\alpha a}}^*B^{j\alpha b}(m_{\chi^0_b}D^{ab}C_1
+m_{\chi^0_a}C^{ab}C_2)\ .
\enea

All the coupling vertices $A$, $B$, $C$, $D$ are defined in the Appendix B.
The $B_{0,1}$ and $C_{0,1,2,00}$ are the standard two-point and three-point
functions with its defination given in the program LoopTools\cite{loop}. 
The arguments of function $B$ from Fig.
(c), (d), (e), (f) are $(M_Z^2,\ m_{\chi^-_a}^2,\ m_{\chi^-_b}^2)$,
$(M_Z^2,\ m_{\chi^0_a}^2,\ m_{\chi^0_b}^2)$, $(0,\ m_{\tilde{\nu}_\alpha}^2,\
m_{\chi^-_a}^2)$ and $(0,\ m_{\tilde{l}_\alpha}^2,\ m_{\chi^0_a}^2)$ respectively.
The arguments of functions $C$ from Figs. (a), (b), (c), (d) are
$(0,\ M_Z^2,\ 0,\ m_{\chi^-_a}^2,\ m_{\tilde{\nu}_\alpha}^2,\ 
m_{\tilde{\nu}_\alpha}^2)$, $(0,\ M_Z^2,\ 0,\ m_{\chi^0_a}^2,\ m_{\tilde{l}_\alpha}^2,\ 
m_{\tilde{l}_\beta}^2)$, $(0,\ M_Z^2,\ 0,\ m_{\tilde{\nu}_\alpha}^2,
\ m_{\chi^-_a}^2,\ m_{\chi^-_b}^2)$, $(0,\ M_Z^2,\ 0,\ m_{\tilde{l}_\alpha}^2,
\ m_{\chi^0_a}^2,\ m_{\chi^0_b}^2)$ respectively, where the external fermion
masses have been set to zero.

\pagebreak

\newpage

\section*{Figure Captions}
\newcounter{FIG}
\begin{list}{{\bf FIG. \arabic{FIG}}}{\usecounter{FIG}}
\item
Feynman diagrams contributing to $\tau\rightarrow\mu\gamma$. 
The universal scalar neutrino
masses become non-degenerate and mixed by the non--diagonal Yukawa couplings.
This effect gives low energy scalar neutrino mixing by RGEs running.
\item
The Feynman diagrams contributing to the process 
$\tau\rightarrow\mu\gamma$.
\item
The Feynman diagrams contributing to the process $Z\rightarrow\mu\tau$.
The other two self-energy diagrams coming from $\tau$ legs are omitted.
\item
Branching ratio of $\tau\rightarrow\mu\gamma$ as a function of $m_0$ for
$m_\frac{1}{2}=100GeV$, $\tan\beta=5$ and $A_0=2m_0,\ m_0,\ 0$. The solid
line is for $\mu\ >\ 0$ and the dashed line is for $\mu\ < \ 0$.
\item
Branching ratio of $\tau\rightarrow\mu\gamma$ as a function of $m_0$ for
$A_0=m_0$, $\tan\beta=5$ and $m_\frac{1}{2}=100GeV,\ 200GeV,\ 400GeV$. 
The solid line is for $\mu\ >\ 0$ and the dashed line is for $\mu\ < \ 0$.
\item
Branching ratio of $\tau\rightarrow\mu\gamma$ as a function of $m_0$ for
$m_\frac{1}{2}=100GeV$, $\tan\beta=10$, $\mu\ >\ 0$ and $A_0=\pm 2m_0,\
\pm m_0,\ 0$. 
The solid line is for $A_0\ > \ 0$ and the dashed line is for $A_0\ < \ 0$.
\item
Branching ratio of $\tau\rightarrow\mu\gamma$ as a function of $m_0$ for
$A_0=m_0$, $m_\frac{1}{2}=100GeV$, and  $\tan\beta=2,\ 5,\ 10$. 
The solid line is for $\mu\ >\ 0$ and the dashed line is for $\mu\ < \ 0$.
\item
Branching ratio of $Z\rightarrow\mu\tau$ as a function of $m_0$ for
$m_\frac{1}{2}=100GeV$, $\tan\beta=5$ and $A_0=2m_0,\ m_0,\ 0$. The solid
line is for $\mu\ >\ 0$ and the dashed line is for $\mu\ < \ 0$.
\item
Branching ratio of $Z\rightarrow\mu\tau$ as a function of $m_0$ for
$A_0=2m_0$, $\tan\beta=5$ and $m_\frac{1}{2}=100GeV,\ 200GeV,\ 400GeV$. 
The solid line is for $\mu\ >\ 0$ and the dashed line is for $\mu\ < \ 0$.
\item
Branching ratio of $Z\rightarrow\mu\tau$ as a function of $m_0$ for
$A_0=2.5m_0$, $m_\frac{1}{2}=150GeV$, and  $\tan\beta=2,\ 5,\ 10$. 
\end{list}

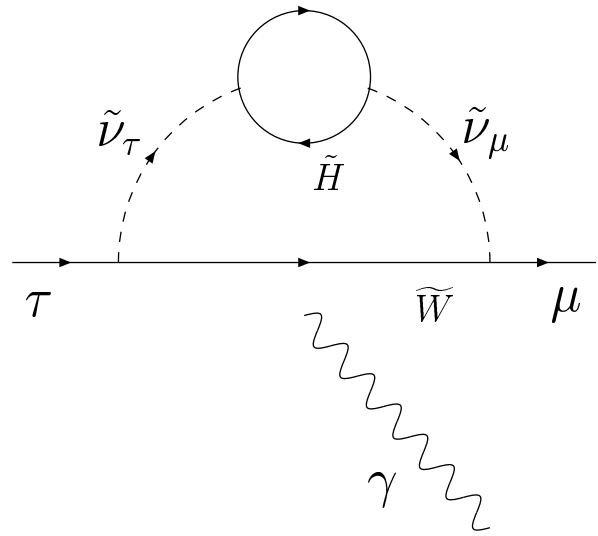
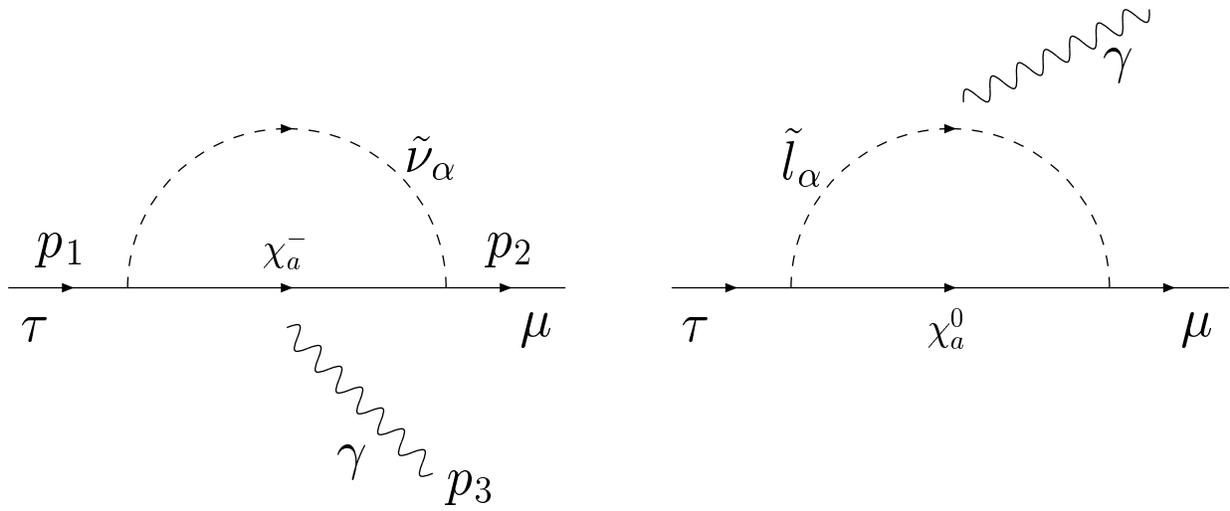
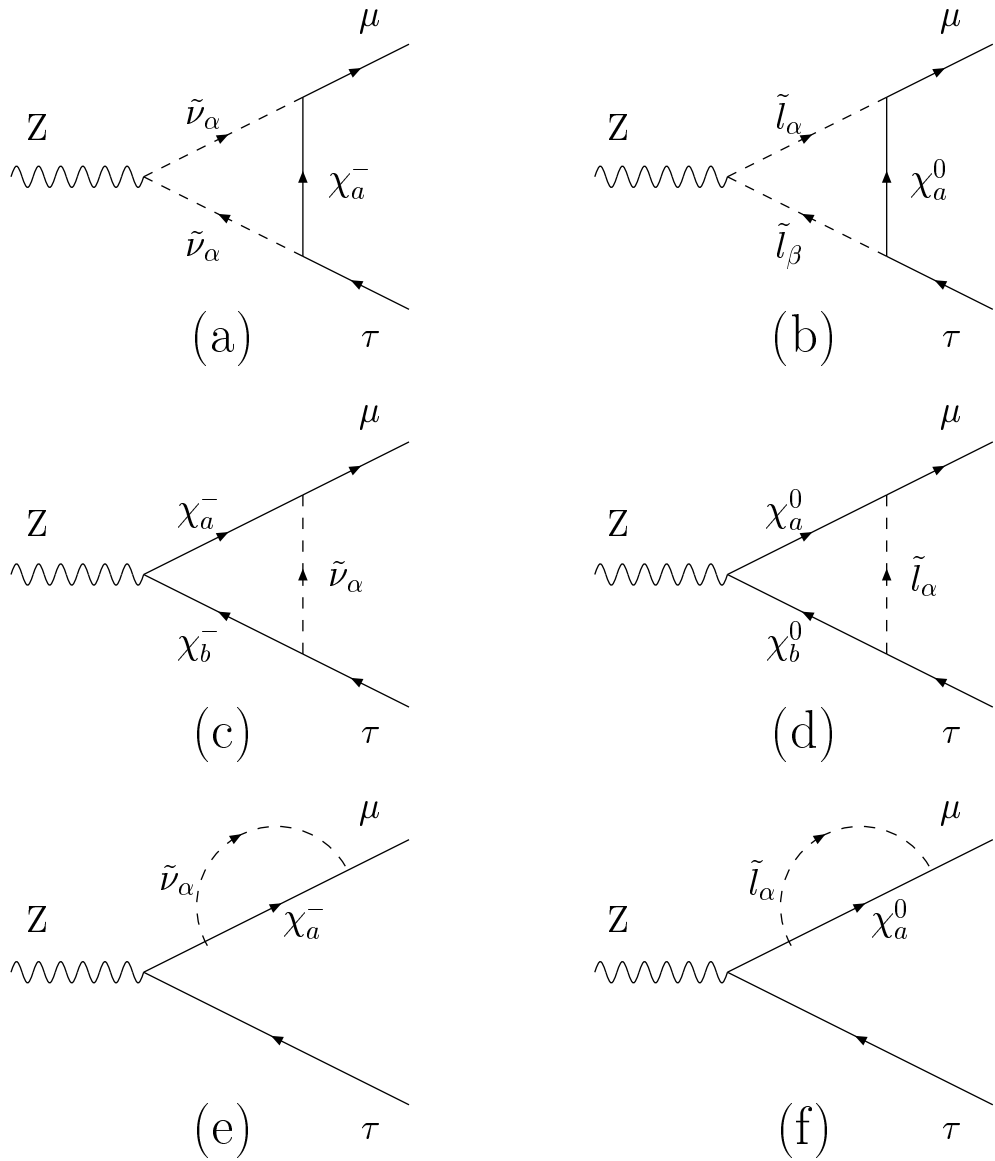
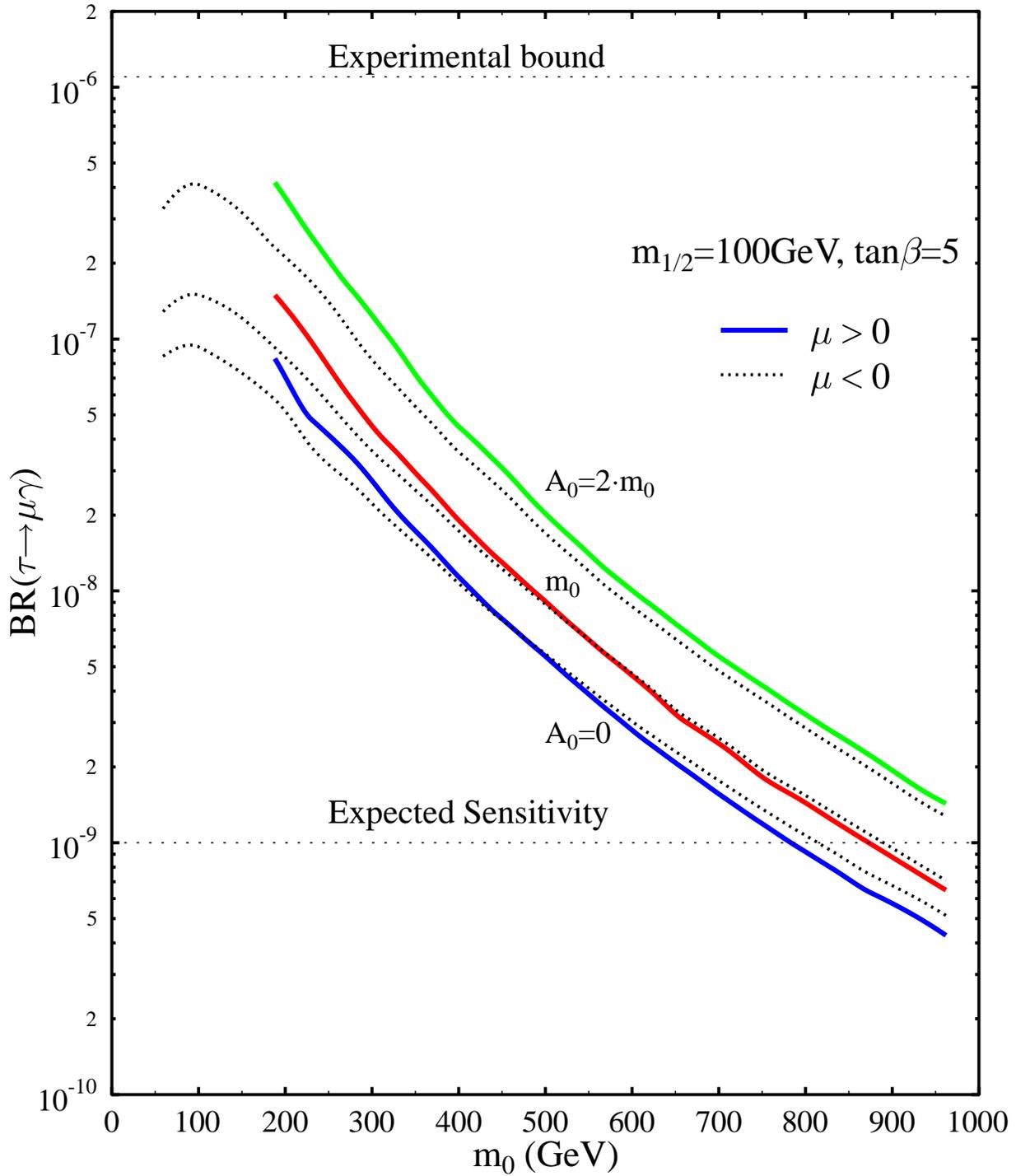
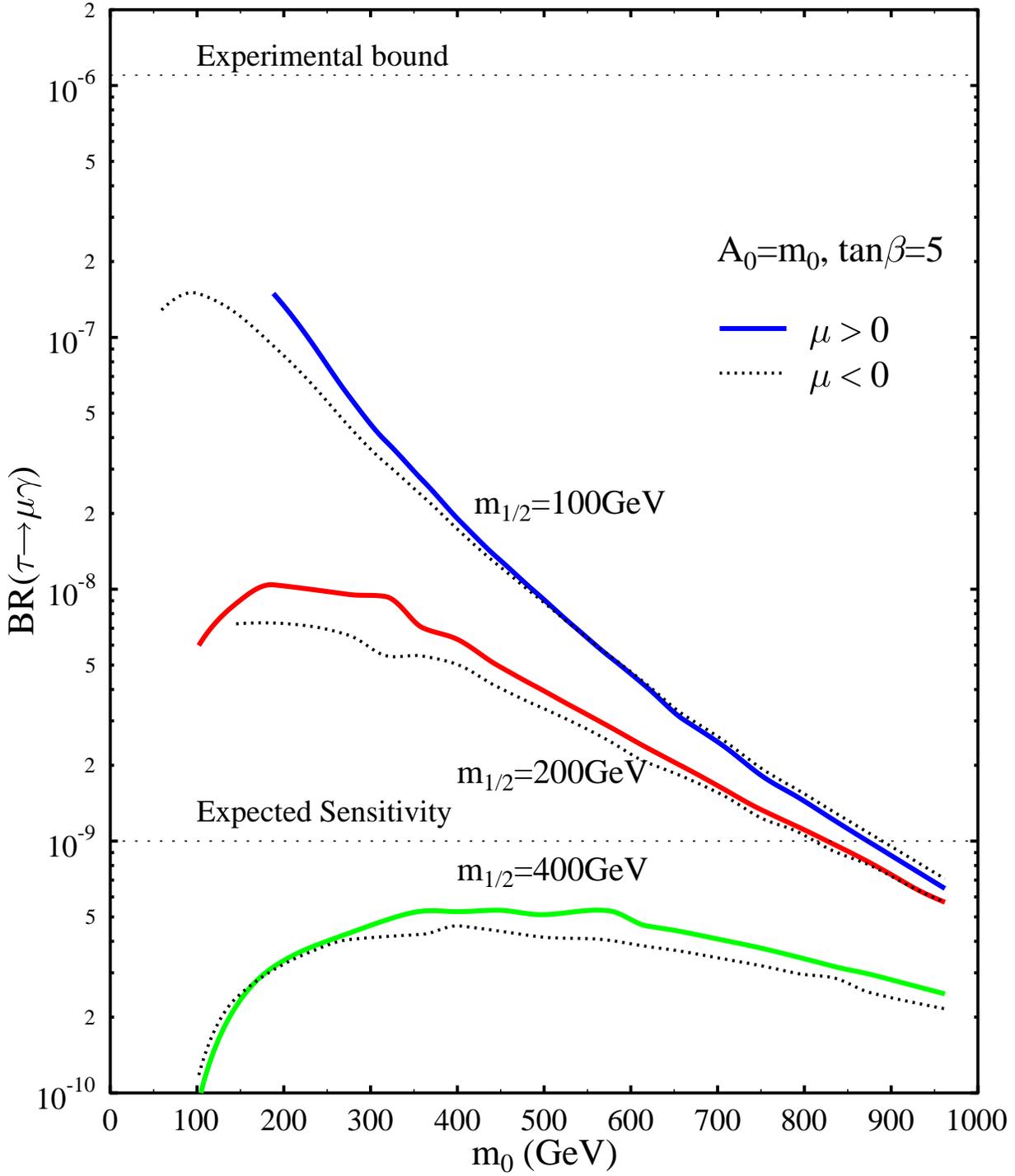
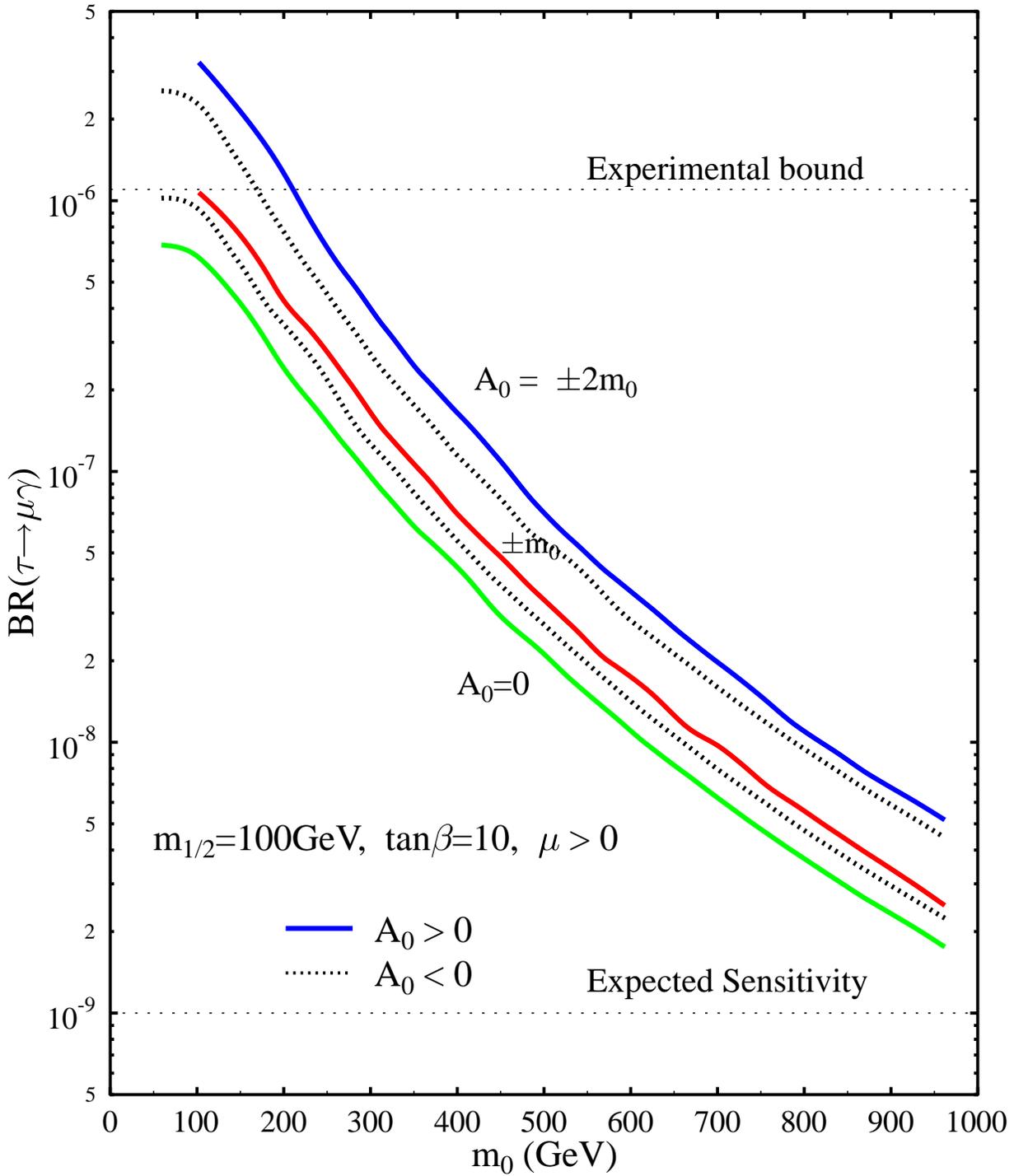
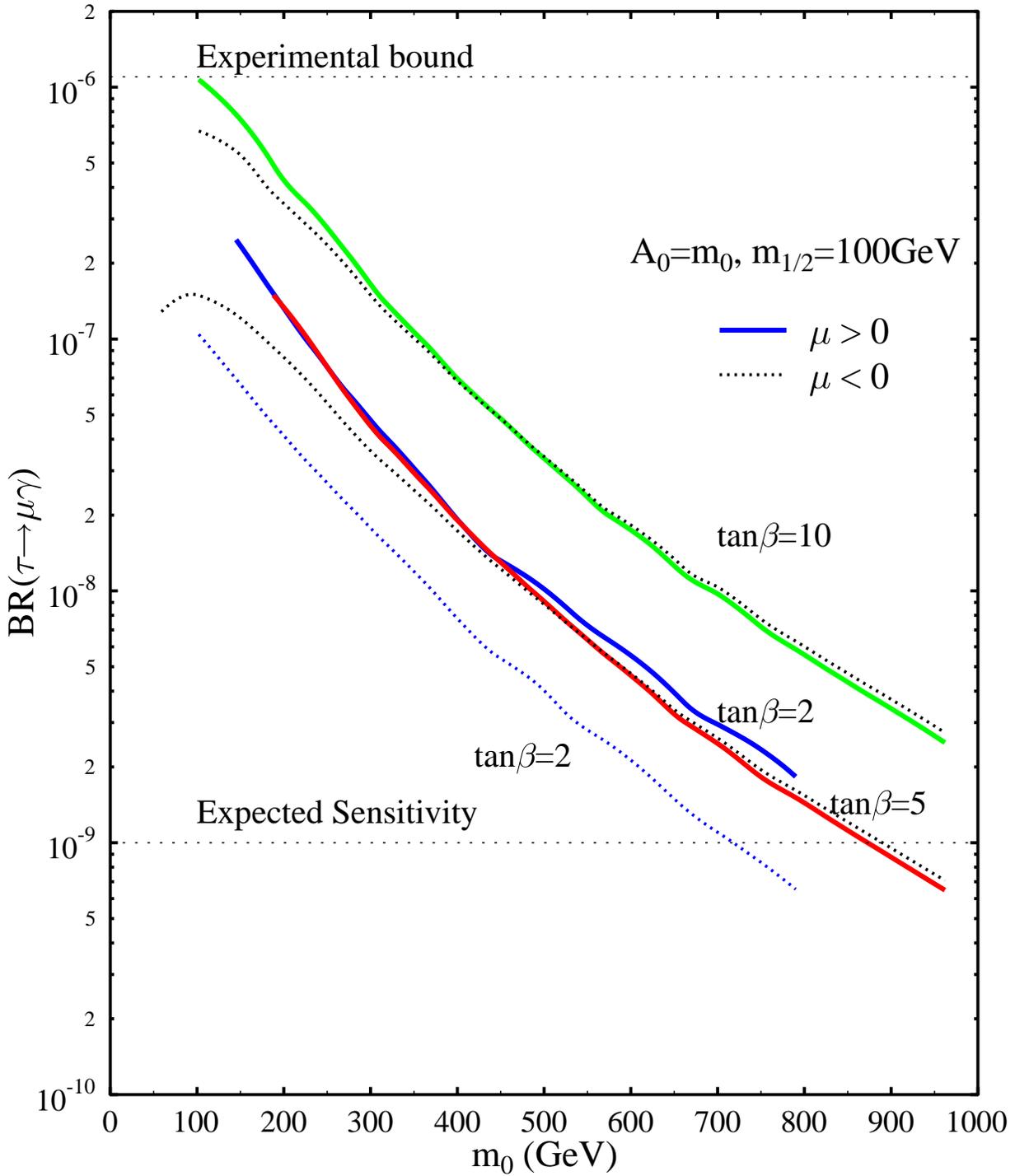
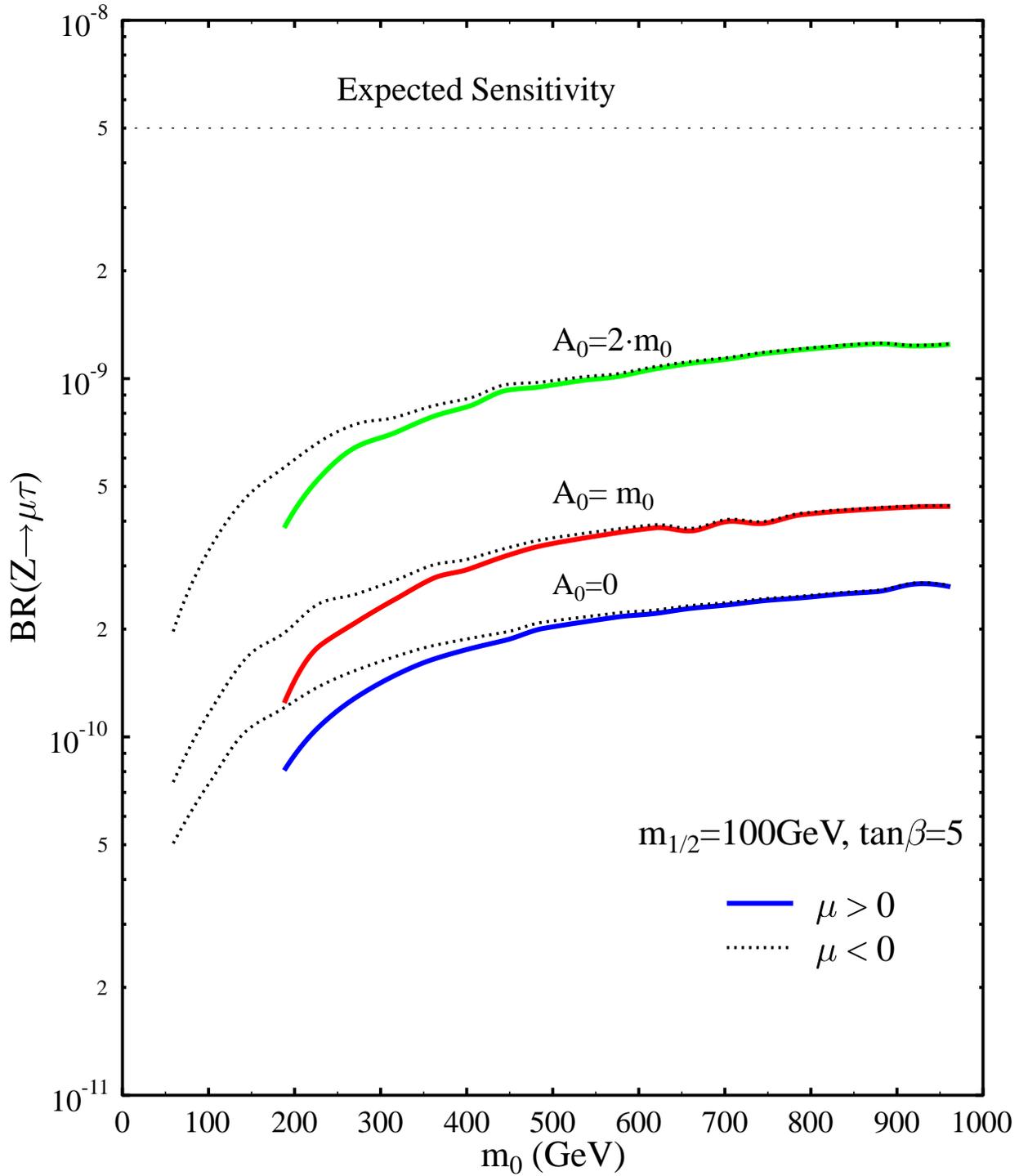
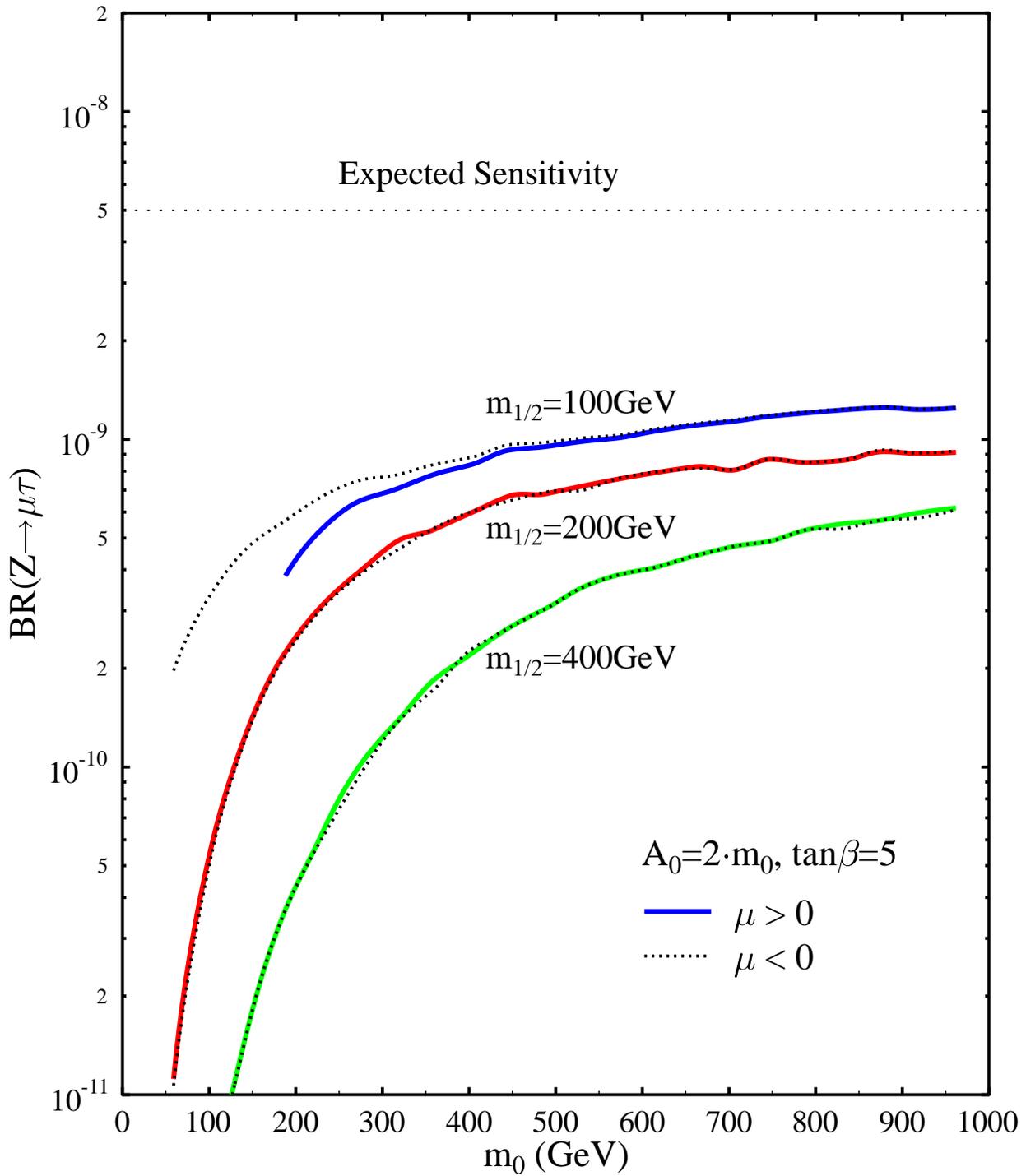
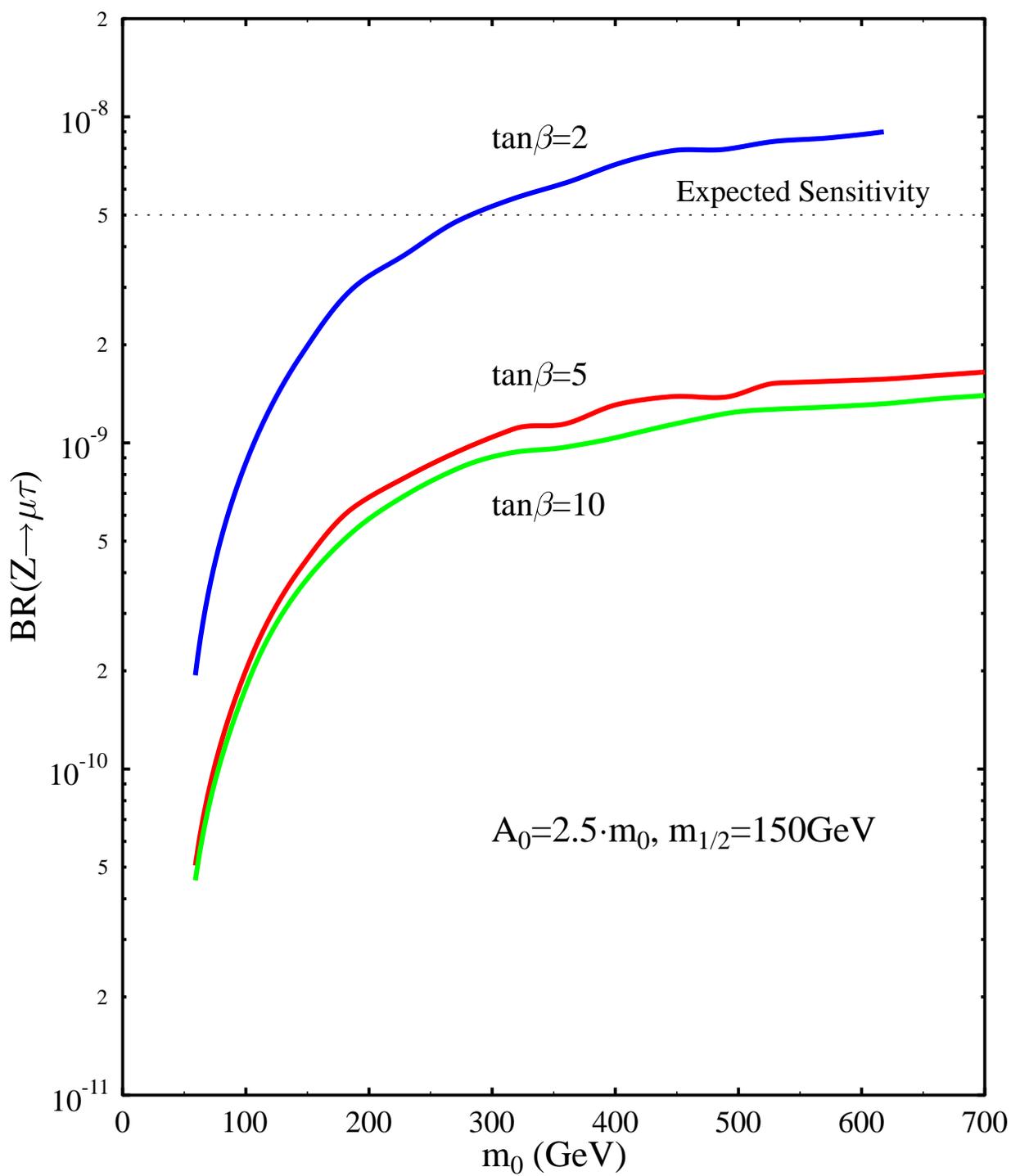
\end{document}